\documentclass[10pt,conference]{IEEEtran}
\usepackage{cite}
\usepackage{amsmath,amssymb,amsfonts}
\usepackage{graphicx}
\usepackage{textcomp}
\usepackage{xcolor}
\def\BibTeX{{\rm B\kern-.05em{\sc i\kern-.025em b}\kern-.08em
    T\kern-.1667em\lower.7ex\hbox{E}\kern-.125emX}}

 \makeatletter
    \newcommand{\linebreakand}{%
      \end{@IEEEauthorhalign}
      \hfill\mbox{}\par
      \mbox{}\hfill\begin{@IEEEauthorhalign}
    }
    \makeatother
    
\IEEEoverridecommandlockouts

\title{Repairing Bugs in Python Assignments Using Large Language Models}

 \author{\IEEEauthorblockN{Jialu Zhang}
    \IEEEauthorblockA{\textit{Yale University} \\
    New Haven, USA \\
    jialu.zhang@yale.edu}
    \and
    \IEEEauthorblockN{Jos\'e Cambronero}
    \IEEEauthorblockA{\textit{Microsoft} \\
    New Haven, USA \\
    jcambronero@microsoft.com}
    \and
    \IEEEauthorblockN{Sumit Gulwani}
    \IEEEauthorblockA{\textit{Microsoft} \\
    Redmond, USA \\
    sumitg@microsoft.com}
    \and
    \IEEEauthorblockN{Vu Le}
    \IEEEauthorblockA{\textit{Microsoft} \\
    Redmond, USA \\
    levu@microsoft.com}
    \linebreakand 
    \IEEEauthorblockN{Ruzica Piskac}
    \IEEEauthorblockA{\textit{Yale University} \\
    New Haven, USA \\
    ruzica.piskac@yale.edu}
    \and
    \IEEEauthorblockN{Gustavo Soares}
    \IEEEauthorblockA{\textit{Microsoft} \\
    Redmond, USA \\
    gsoares@microsoft.com}
    \and
    \IEEEauthorblockN{Gust Verbruggen}
    \IEEEauthorblockA{\textit{Microsoft} \\
    Keerbergen, Belgium \\
    gverbruggen@microsoft.com}\thanks{This work was performed when Jialu Zhang was an intern at Microsoft Prose Team. All other authors are listed in alphabetical order.}
    }

\usepackage{xspace}
\usepackage{footnote}
\usepackage{hhline}
\usepackage{diagbox}
\usepackage{multirow}
\usepackage{setspace}
\usepackage{epsfig}
\usepackage{url}

\usepackage{breakurl}
\usepackage{booktabs}
\usepackage{hyperref}

\usepackage{xcolor}
\usepackage{lipsum}
\usepackage{courier}
\usepackage{listings}
\usepackage{caption}
\usepackage{colortbl}
\usepackage{arydshln} 
\usepackage{makecell} 
\usepackage{tikz}
\usepackage{color}
\usepackage{rotating}
\usepackage{subcaption}

\usepackage{varwidth}
\usetikzlibrary{fit,arrows,calc,positioning,quotes}

\usepackage{pgfplots}

\usepackage{mathtools}
\usepackage{fancyvrb}
\usepackage{textcomp}
\usepackage{microtype}

\usepackage{stmaryrd}

\usepackage{algorithm}
\usepackage[noend]{algpseudocode}

\usepackage{alltt}

\tikzset{
  -|-/.style={
    to path={
      (\tikztostart) -| ($(\tikztostart)!#1!(\tikztotarget)$) |- (\tikztotarget)
      \tikztonodes
    }
  },
  -|-/.default=0.5,
  |-|/.style={
    to path={
      (\tikztostart) |- ($(\tikztostart)!#1!(\tikztotarget)$) -| (\tikztotarget)
      \tikztonodes
    }
  },
  |-|/.default=0.5
}

\usetikzlibrary{shapes.geometric}

\long\def\com#1{}

\newcommand{\app}{\texttt{MMAPR}\xspace}
\newcommand{\system}{\texttt{MMAPR}\xspace}

\newcommand{\para}[1]{\smallskip\noindent {\bf #1}}

\newcommand{\squishlist}{
   \begin{list}{$\bullet$}
    { \setlength{\itemsep}{0pt}      \setlength{\parsep}{3pt}
      \setlength{\topsep}{3pt}       \setlength{\partopsep}{0pt}
      \setlength{\leftmargin}{3.5mm} \setlength{\labelwidth}{1em}
      \setlength{\labelsep}{0.5em} }
}

\newcommand{\squishend}{
    \end{list}  }

\def\BibTeX{{\rm B\kern-.05em{\sc i\kern-.025em b}\kern-.08em
    T\kern-.1667em\lower.7ex\hbox{E}\kern-.125emX}}

\begin{document}

\maketitle

\begin{abstract}
Students often make mistakes on their 
introductory programming assignments as part of their learning process.
Unfortunately, providing custom repairs for these mistakes can require a 
substantial amount of
time and effort from class instructors. Automated program repair (APR) techniques can be
used to synthesize such fixes. Prior work has explored the use of symbolic and
neural techniques for APR in the education domain. Both types of approaches 
require either substantial engineering efforts or
large amounts of data and training. We propose to use a large language model
trained on code, such as Codex, to build an APR system -- \system{} -- for
introductory Python programming assignments. Our system can fix
\emph{both} syntactic and semantic mistakes by combining multi-modal
prompts, iterative querying, test-case-based selection of few-shots,
and program chunking. We evaluate \system{} on 286 real student programs and
compare to a baseline built by combining a state-of-the-art Python 
syntax repair engine, BIFI, and state-of-the-art Python semantic repair
engine for student assignments, Refactory. We find that \system{} can fix
more programs and produce smaller patches on average.
\end{abstract}

\section{Introduction}
\label{sec:intro}

Programming education has grown substantially
in popularity in the past decade~\cite{rise-of-cs}. 
A key challenge associated with this growth is 
the need to provide 
novice students with efficient and effective learning support. In an ideal world, 
teaching assistants
would monitor students' learning process, and when students' code is not correct, they would then
help them to derive a correct solution. 
However, this approach does not scale and educational institutions 
struggle to find teaching assistants. As a result,
there is an interest in developing automated tools that students can use for feedback instead. These 
tools provide custom repairs for their programming mistakes. 
The field of automated program repair (APR),
which has a long history in the software engineering
community~\cite{GenProg, genesis, prophet, Angelix, verifix, legoues-cacm2019}, has introduced 
different approaches~\cite{SKP, clara, rolim2017learning, Refactory} 
to produce such automated repairs for
student mistakes in introductory assignments. 
Given a buggy student program, the APR system
aims to produce a patch that satisfies a specification (typically
the instructor-provided test cases). The patch
must also minimize the number of changes made, with the goal of facilitating student learning~\cite{Refactory}.

Prior automated program repair systems
for student programming assignments have generally been implemented using
purely symbolic~\cite{clara, SAR, Refactory, rolim2017learning}
or 
purely neural~\cite{SKP, ahmed2018compilation} techniques. 
Symbolic approaches require substantial engineering efforts to develop, typically
requiring significant program analysis/repair experience, as well as
custom repair strategies tailored to the language domain in which students implement
their assignments. Neural approaches mitigate some of the engineering challenges but  
typically require substantial amounts of data, often leading to
specialized use cases for Massive Open Online Courses (MOOCs). Furthermore,
these systems are typically tailored to focus exclusively on syntax repair or exclusively on semantic repair. 
For the latter, the assumption is the code to be repaired contains no syntactic errors.

In this paper, we introduce \system{}, a \textbf{\texttt{m}}ulti-\textbf{\texttt{m}}odal \textbf{\texttt{a}}utomated
\textbf{\texttt{r}}epair \textbf{\texttt{s}}ystem. \system{} is
a \emph{unified} syntax and semantic repair engine
for introductory Python programming assignments. 
We use a
large language model trained on code (LLMC)
as the core component in \system{}. Using an LLMC 
removes the need for custom symbolic repair logic or 
retraining of a new neural model, and it allows us to handle \emph{both}
syntactic and semantic mistakes. While LLMCs have been successfully applied
to tasks such as code generation~\cite{copilot}, its impact in the education domain
remains controversial~\cite{coping-with-copilot}. 
Using an LLMC for repair provides an opportunity to produce a positive impact in
this domain.

We follow the approach
of recent work~\cite{xia2022less, joshi2022repair} in framing
program repair as a code generation task that can be tackled
with an LLMC.  However, using LLMCs to produce student repairs requires 
addressing three challenges.
First, in the classroom setting, multiple sources can help to repair students' code. 
For example, the instructor might provide test cases that the solution code needs to pass,
and/or a description
of the task in a natural language. Even the compiler messages can be used to repair
syntax errors in the code. While standard LLMC-based APR tools take as input
the buggy code and only one correctness criterion (for example, test cases), 
our approach is multi-modal. In addition to the student's buggy program, 
we also take all the above-mentioned sources as a part of the input. Moreover,
we make effective use of
other students' submissions, if they are available.
Second, we need to mitigate the extent to which the LLMC can generate more
code than necessary or make changes to parts
of the program that are not incorrect, which could result
in an excessive patch.
Third,
using an LLMC as a black box means that we also need to adapt traditional prompt 
engineering techniques. We use
\emph{prompt-based learning}~\cite{liu2021pre}, which consists
of creating a text-based template to generate the code.

\system{} 
\emph{ensembles} multi-modal prompts to generate complementary
repair candidates. It employs prompts in an iterative
querying strategy that first uses syntax-targeted prompts
and \emph{then} semantics-targeted prompts. 
\system{}  takes inspiration from 
existing symbolic repair 
literature~\cite{ke2015repairing, SAR, clara} 
and
leverages few-shot learning, which adds task-related
examples to the prompt, by retrieving other student's programs that 
have similar mistakes (and eventual corrections).
To identify these programs, \system{} computes a similarity metric
over test-suite outcomes. Finally, to reduce the number of changes induced by syntax errors that should have relatively simple fixes, \system{} uses the program's structure to extract a subprogram to 
give as input to the LLMC. By reducing the code surface exposed to the LLMC, \system{} biases repairs towards
fewer edits.

We evaluated \system{} on student programs, from an introductory
Python programming course at a major university in India.
Our evaluation has 15 programming tasks, totalling 286
student programs. These student programs contain both syntactic and semantic mistakes. As there is
currently no tool that can solve both errors simultaneously, we combined BIFI~\cite{BIFI}
with Refactory~\cite{Refactory} to create a state-of-the-art baseline system.
BIFI is a state-of-the-art transformer-based
syntax repair engine for Python, while Refactory is a state-of-the-art symbolic
semantic repair engine for introductory Python assignments.

Our results show that \system{}
can effectively repair student programs
in our benchmark set. While the baseline
repairs 67.13\% of programs, \system{} (without few-shot learning) can repair 86.71\%. If we add few-shot learning to 
\system{}, this repair rate rises to 96.5\%.
Furthermore, the average token edit
distance associated with \system{}
patches are smaller (31.4 without few-shots and 31.29 with few-shots) compared to
the patches produced by the baseline (42.50).


We carried out an ablation study
to understand the impact of
our design decisions. 
Our results indicate 
that by performing  iterative querying the repair
rate rises from 82.87\% to 86.71\%.
Furthermore, adding few-shots raises the repair success rate
to 96.5\%. The evaluation also shows that our techniques are
important for maintaining the repaired program  
similar to the buggy input program. For example, removing the program chunker,
which selects subprograms in the
syntax repair phase, raises the average
token edit distance from 5.46 to 9.38.
We also show that
different multi-modal prompts have varying
performance, but if we combine their
candidates  as 
we do in \system{}, we obtain the best
performance.

To summarize, we make the following contributions:
\begin{itemize}
    \item  We propose an approach to automatically repair mistakes in students' Python programming assignments
    using a large language model trained on code (LLMC). Our approach
    uses multimodal prompts, iterative querying, test-case-based few-shot selection,
    and structure-based program chunking to repair student mistakes.  In contrast to prior work,
    our approach uses the \emph{same} underlying LLMC to repair \emph{both} syntactic
    and semantic mistakes.

    \item We implement this approach in \system{}, which uses OpenAI's popular Codex as the LLMC. We evaluate \system{} on a dataset of 286 real student Python programs
    drawn from an introductory Python programming course in India. We compare performance to a baseline produced by combining a state-of-the-art syntax repair engine (BIFI)
    and a state-of-the-art semantic repair engine (Refactory). Our results show that 
    \system{} can outperform our baseline, even without few-shot learning. 
    In addition, on average, \system{}'s
    patches are closer to the original
    student submission.
\end{itemize}

The remainder of the paper is structured as follows.
Section~\ref{sec:motiv} walks through multiple examples of
real student mistakes, as well as the associated \system{} patches. 
Section~\ref{sec:background} provides a brief background on concepts related
to large language models. Section~\ref{sec:method} describes our
approach in detail and its implementation in \system{}. 
Section~\ref{sec:eval} provides experimental results on our dataset of student Python programs,
including a comparison to a baseline repair system. We discuss related work in Section~\ref{sec:related}.
Finally, we conclude with takeaways in Section~\ref{sec:conclusion}.
\section{Motivating Example}
\label{sec:motiv}

Consider Figure~\ref{fig:motiv1}, which shows
a student's incorrect program, along with
a solution generated by \system{}.
The student is solving the task of reading
two numbers from stdin and printing
different things depending on whether both,
either, or neither of the values are prime.

The student has made both syntactic and semantic
mistakes. Lines 1 and 2  call \lstinline|input|
twice to read from stdin, and parse these values
as integers using \lstinline|int|. However, this constitutes a
semantic mistake, as the assignment input format
consists of two values \emph{on the same line}
separated by a comma. Furthermore, a
traditional semantic repair engine would fail
to fix this student's assignment as there is
\emph{also} a syntactic mistake at line 30. The
student used a single \lstinline|=| for comparison
in the \lstinline|elif| clause (the correct syntax
would be a double equals).

The \system{} solution, shown alongside it,
fixes the input processing (semantic mistake)
by reading from stdin, splitting on the comma,
and applying \lstinline|int| (to parse as integer)
using the \lstinline|map| combinator. Line 23 fixes
the syntax error by replacing single equals with
double equals (for comparison). 
Interestingly, the underlying LLMC (Codex) \emph{also}
refactored the student's program. In this case, 
lines 8 through 17 correspond to a function to 
check if a number is prime. This function is called
twice, at lines 18 and 19. This replaces the
repeated code in the original program, which 
spanned lines 9-17 and lines 17-26.

The edit distance between the \system{}
repair and the original student program is
95, while the distance between
the instructor's reference solution and the original
student program is 188. A smaller edit distance
is a key goal for APR in the educational domain,
as this can help the student understand the
repair with respect to their own mistakes.

\begin{figure}[t]
    \centering
    \begin{subfigure}[b]{.5\columnwidth}
    \lstset{
    language=Python,
    numbers=left,
    stepnumber=1,
    escapeinside={<@}{@>}
    }
	\begin{lstlisting}
<@\textcolor{red}{a=int(input())}@>
<@\textcolor{red}{b=int(input())}@>
n=str(a)
m=str(b)
n1=n[::-1]
m1=m[::-1]
x=int(n1)
y=int(m1)
if x>1:
  for i in range(2, x):
    if (x%i)==0:
      p=0
        break
        else:
          p=1
else:
  p=0
if y>1:
  for j in range(2, y):
    if (y%j)==0:
      q=0
        break
      else:
        q=1
else:
  q=0
if p==1 and q==1:
  g=x+y
  print(g)
<@\textcolor{red}{elif p=1 or q=1:}@>
  h=a+b
  print(h)
else:
  print(a*b)
\end{lstlisting}
\caption{Incorrect Program}
\label{subfig:motiv1_incorrect}
    \end{subfigure}%
\hfill
        \begin{subfigure}[b]{.45\columnwidth}
    \lstset{
    language=Python,
    numbers=left,
    stepnumber=1,
    escapeinside={<@}{@>}
    }
	\begin{lstlisting}
<@\textcolor{blue}{a,b = map(int, input().split(','))}@>
n=str(a)
m=str(b)
n1=n[::-1]
m1=m[::-1]
x=int(n1)
y=int(m1)
<@\textcolor{blue}{def prime(z):}@>
  <@\textcolor{blue}{if z>1:}@>
    <@\textcolor{blue}{for i in range(2,z):}@>
      <@\textcolor{blue}{if (z\%i)==0:}@>
        <@\textcolor{blue}{return 0}@>
          <@\textcolor{blue}{break}@>
      <@\textcolor{blue}{else:}@>
        <@\textcolor{blue}{return 1}@>
    <@\textcolor{blue}{else:}@>
      <@\textcolor{blue}{return 0}@>
p=prime(x)
q=prime(y)
if p==1 and q==1:
  g=x+y
  print(g)
<@\textcolor{blue}{elif p==1 or q==1:}@>
  h=a+b
  print(h)
else:
  print(a*b)
\end{lstlisting}
\caption{\app's Repair}
\label{subfig:motiv1_correct}
    \end{subfigure}%
    \caption{
    A student's submission contains both syntax and semantic
    mistakes (red). \system{}'s fixes (blue) the original
    semantic and syntactic issues and also refactors part of the student's code into a function (lines 8 - 17 in (b))
    that avoids code duplication (lines 9-17, 18-26 in (a)).
    }
    \label{fig:motiv1}
\end{figure}


Figure~\ref{fig:motiv2} presents
another example of an incorrect student program and
a solution generated by \system{}. In this assignment, the students need to check whether a string, read from stdin, is a palindrome or not, and
print out a message accordingly to stdout.
For this student's program, \system{} has to generate a complex repair
that fixes four syntax mistakes and multiple semantic bugs.

The student has made syntax errors on lines 4, 8, 10, and 12, where
they have left off the colon symbol necessary for control flow statements
in Python. On line 2, the student called a non-existent function
\lstinline|lower|. The student has used standard division
on lines 5, 6, 13, and 14 when they should have used
integer division. The student has included two spurious print 
statements, at lines 7 and 15, which will interfere with
the instructor's test-suite execution, as the suite checks 
values printed to stdout for correctness. Finally, the student
has omitted the expected print statements (along with the equality
check) for the case where the input string is of even length.

While the student's program has many mistakes, the overall structure
and key concepts are there. Looking at the \system{} solution
shown alongside, it resolves these mistakes but preserves
the student's overall structure. In particular, \system{}
replaces the non-existent \lstinline|lower| function with
a call to the string method with the same name.
It replaces the division operator (/) throughout the program
with the intended floor division operator (//), comments
out the extract print statements, and adds the missing equality
check and print statements in the case of even-length inputs.

The edit distance between the \system{}
repair and the original student program is
52, while the distance between
the instructor's reference solution and the original
student program is 97. The reference solution is a standard
one line program for palindrome. 
Once again, the \system{} repair is closer to the student submission than
the instructor's reference solution.

\begin{figure}[t]
    \centering
    \begin{subfigure}[b]{.5\columnwidth}
    \lstset{
    language=Python,
    numbers=left,
    stepnumber=1,
    escapeinside={<@}{@>}
    }	
\begin{lstlisting}
i = input()
<@\textcolor{red}{S = lower(i)}@>
l = len(S)
<@\textcolor{red}{if(l\%2!=0)}@>
  B = <@\textcolor{red}{S[:(l+1)/2]}@>
  E = <@\textcolor{red}{S[:(l+1)/2:-1]}@>
  <@\textcolor{red}{print(B,E)}@>
  <@\textcolor{red}{if(B==E)}@>
    print(i,'is a palindrome.')
  <@\textcolor{red}{else}@>
    print(i,'is NOT a palindrome.')
<@\textcolor{red}{else}@>
  B = <@\textcolor{red}{S[:l/2]}@>
  E = <@\textcolor{red}{S[:l/2:-1]}@>
  <@\textcolor{red}{print(B,E)}@>
\end{lstlisting}
\caption{Incorrect Program}
\label{subfig:motiv2_incorrect}
    \end{subfigure}%
\hfill
        \begin{subfigure}[b]{.45\columnwidth}
    \lstset{
    language=Python,
    numbers=left,
    stepnumber=1,
    escapeinside={<@}{@>}
    }
	\begin{lstlisting}
i = input()
<@\textcolor{blue}{S = i.lower()}@>
l = len(S)
if(l%2!=0)<@\textcolor{blue}{:}@>
  <@\textcolor{blue}{B = S[:(l+1)//2]}@>
  <@\textcolor{blue}{E = S[:l//2-1:-1]}@>
  <@\textcolor{blue}{\#print(B,E)}@>
  if(B==E)<@\textcolor{blue}{:}@>
    print(i,'is a palindrome.')
  else<@\textcolor{blue}{:}@>
    print(i,'is NOT a palindrome.')
else<@\textcolor{blue}{:}@>
  <@\textcolor{blue}{B = S[:l//2]}@>
  <@\textcolor{blue}{E = S[:l//2-1:-1]}@>
  <@\textcolor{blue}{\#print(B,E)}@>
  <@\textcolor{blue}{if(B==E):}@>
    <@\textcolor{blue}{print(i,'is a palindrome.')}@>
  <@\textcolor{blue}{else:}@>
    <@\textcolor{blue}{print(i,'is NOT a palindrome.')}@>
\end{lstlisting}
\caption{\app's Repair}
\label{subfig:motiv2_incorrect}
    \end{subfigure}%
    \caption{
    A complex repair (blue) that fixes multiple syntax and semantic mistakes (red). The repair produced by 
    \system{}, which preserves the overall structure of the student's program, makes fewer changes to the student's program than a patch with respect to the
    instructor's reference solution.
    }
    \label{fig:motiv2}
\end{figure}

\section{Background}
\label{sec:background}

We now provide a short background on 
concepts related to large language models.

A {\emph{large language model}} (LLM) can be viewed as a probability distribution over sequences of words. This distribution is learned using a deep neural network with a large number of parameters. These networks are typically
trained on large amounts of text (or code) with objectives such
as predicting particular masked-out tokens or \emph{autoregressive}
objectives such as predicting the next token given the 
preceding tokens. When the LLM has been trained
on significant amounts of code, we refer to it
as a large language model trained on code (LLMC).

Often, LLMs are pre-trained and then \emph{fine-tuned}, meaning trained further on more specialized data or tasks. A particularly popular LLMC
is OpenAI's Codex\cite{codex}, a variant of GPT-3~\cite{gpt3} that is fine-tuned on code from more than 50 million GitHub repositories.

In contrast to traditional supervised machine learning,
LLMs have shown to be effective for  \emph{few-} and even \emph{zero-}shot
learning. This means that the LLM can perform tasks it was
not explicitly trained for just by giving it a few examples of the
task or even no examples, respectively, at inference time.

In this setting of few- (or zero-)shot learning, the LLM
is typically employed using what is termed
\emph{prompt-based learning}~\cite{liu2021pre}. 
A \emph{prompt} is a textual template that can be
given as input to the LLM to obtain 
a sequence of iteratively predicted next tokens, called a \emph{generation}. A prompt typically consists of
a query and possibly zero or more examples of the task, called shots. 
For example, the prompt below includes
a specific query to fix a syntax error.
One valid generation, that fixes
the syntax error, would be \lstinline|print()|.

\begin{lstlisting}[]
# Fix the syntax error of the program #

# Buggy program #
print("\")
\end{lstlisting}

In practice, a prompt can
incorporate anything that can be captured in textual format.
In particular, \emph{multi-modal} prompts are those that
incorporate different \emph{modalities} of inputs, such
as natural language, code, and data.

Different prompts may result in different LLM completions.
Other factors may also affect the completions produced, such
as the sampling strategy or hyperparameters for the sampling
strategy. One important hyperparameter is \emph{temperature}, which controls the extent to which we sample less likely completions. 

While we use OpenAI's Codex in this work, there are
other such LLMs that could be used such as
Salesforce's CodeGen~\cite{CodeGen} or OpenScience's BLOOM~\cite{BLOOM}. 
Even within OpenAI's Codex there
are different underlying models offered, 
including Codex-Edit~\cite{codex-edit-api}.
We found performance to be better with the standard Codex completion model.
We now leverage these concepts to describe our approach to APR.
\section{Methodology}
\label{sec:method}

\begin{figure*}
    \centering
    \includegraphics[scale=0.5]{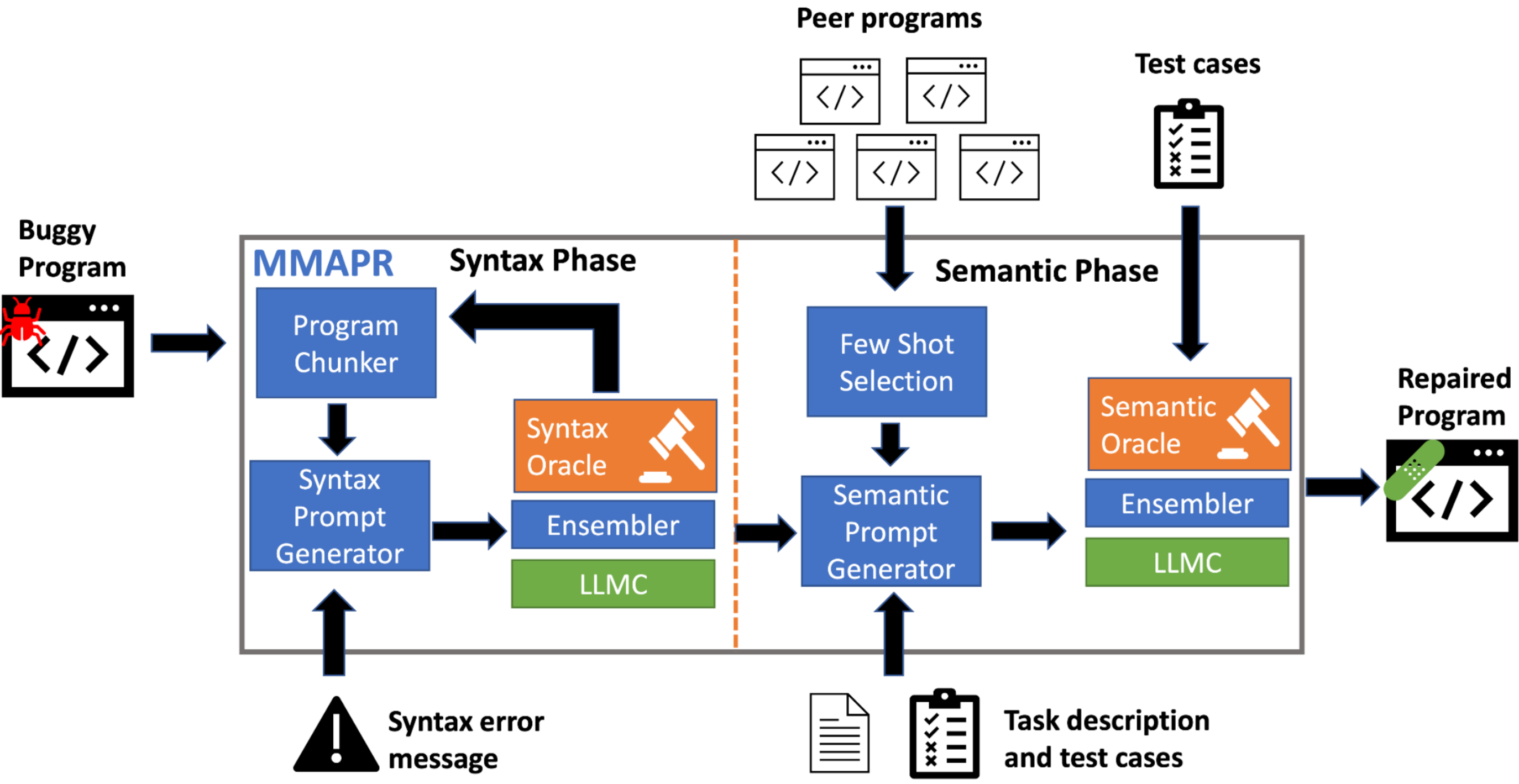}
    \caption{\system{} architecture. 
    A buggy program first enters a syntax repair phase. In this phase, \system{}
    transforms the program using a program chunker, which performs a structure-based
    subsetting of code lines to narrow the focus for the LLMC. Multiple syntax-oriented prompts are generated using this subprogram, fed to an LLMC, and any patches are integrated into the original program. If any candidate satisfies the syntax oracle, it can move on to the semantic phase. In the semantic phase, \system{} leverages both the natural language description of the assignment and the instructor-provided test cases to create various prompts. In addition, if available, \system{} can use other peers' solutions as few-shots by selecting them using test-case-based selection to identify failures that resemble the current student's program, along with eventually correct solutions. Prompts are fed to the LLMC to generate candidates. If multiple candidates satisfy the test suite, \system{} returns the one
    with the smallest edit distance with respect to the original student program.
    }
    \label{fig:system-arch}
\end{figure*}

Figure~\ref{fig:system-arch} provides an overview of the
architecture underlying \system{}. The student's buggy 
program first enters a syntax repair phase. In this phase,
we extract subprograms from the original program that
have a syntax error. Each such subprogram is fed to 
a syntax prompt generator that produces multiple syntax-oriented
prompts. The LLMC then generates repair candidates, which are validated by the syntax oracle. This process is repeated until
all syntax errors are removed. Any candidate that has
no syntax errors moves on to the semantic phase.
In this phase, \system{} uses a semantic prompt generator
to produce semantics-oriented prompts. If it has access
to other student's assignment history, \system{} can also
add few-shots to these prompts. These prompts are
fed to the LLMC, which generates new program candidates.
These are validated by the test-suite-based semantic oracle.
If multiple candidates satisfy all test cases, \system{}
returns the one with the smallest token edit distance with
respect to the student's original buggy program.
We now describe each
step in detail.

\subsection{Syntax Phase}
Students typically first resolve syntax errors in their
assignments, and then move on to resolve semantic errors (such
as test case failures). \system{} takes inspiration
from this approach and similarly splits its repair into
syntax and semantic phases.

In the first phase, \system{} receives the student's buggy program.
A syntax oracle, for example, the underlying Python parser, is used to determine
if there is a syntactic mistake. If there is no such mistake, the program can move into the
semantic phase. However, if there is a mistake, \system{} must produce a patch that resolves it, before moving to the semantic phase.

While our syntax prompt generator could directly include
the original program in the prompt, we have found that doing so
can result in spurious edits that are not actually necessary
to resolve the syntax error. Existing work has also
observed similar phenomena in the related area of natural
language to code generation~\cite{synchromesh}. As a result, we introduced
a component we call the \emph{program chunker} to mitigate this challenge
by reducing the amount of code included in the prompt.

\subsubsection{Program Chunking}

For each syntax mistake in the original buggy program, the program chunker extracts
a subset of lines that contains (1) the oracle-reported syntax error location and (2) the nearest encompassing control-flow statement.
These \emph{chunks} are a heuristic approximation of a basic block, and allow us to restrict
the code input given to the LLMC. Note that we perform this heuristic approximation
as a standard analysis to extract basic blocks typically requires
a syntactically correct input program.

\begin{algorithm}
\begin{flushleft}
        \textbf{Input: }$\textit{sC}$ : Program Source Code \\
        \textbf{Input: }$\textit{msg}$ : Compiler Message \\
        \textbf{Output: }$\textit{chunkedCode}$ : Chunked Program Source Code \\
\end{flushleft}
\begin{algorithmic}[1]
{\small 
\Procedure{chunker}{$sC,msg$}
  \State $\textit{listsC,errorLine}$ = locateError($sC,msg$)
  \State $\textit{indentLevel}$ = getIndentationLevel($listsC, errorLine$)
  \\
  \Comment{slice adjacent code with no less than the indentation level}
  \State $\textit{start, end}$ = sliceBiway($listsC, errorLine, indentLevel$) 
  \\
  \Comment{adjust slice pointers to account for containing control-flow}
  \If {$listsC[start]$.startswith(cfNodes)}
    \State $\textit{newindentLevel}$ = getIndentationLevel($listsC, start$)
    \State $\textit{start, end}$ = sliceBiway($listsC, start, newindentLevel$)
  \EndIf
  \State \Return $\textit{chunkedCode}$ = listsC[start,end]
\EndProcedure

}
\end{algorithmic}
\caption{Chunker: slicing the program snippet that contains the error message}
\label{algo:chunker}
\end{algorithm}

\system{} extracts the program chunk for the first (top-down) syntax error
reported. Algorithm~\ref{algo:chunker} outlines the procedure used to produce
this program chunk. It takes advantage of both control-flow structure
(based on Python keywords) and indentation, which is meaningful in the
Python language.
The program chunker first identifies the adjacent code that has the same or large indentation level as the line with the syntax error.
Then, if the code chunk contains control flow related keywords, such as \texttt{if} and \texttt{elif}, \system{} makes sure the corresponding keywords (such as \texttt{elif} or \texttt{else}) is also in the chunk.
This program chunk is then provided to the syntax prompt generator.

\subsubsection{Syntax Prompt Generator}
\begin{figure}[t]
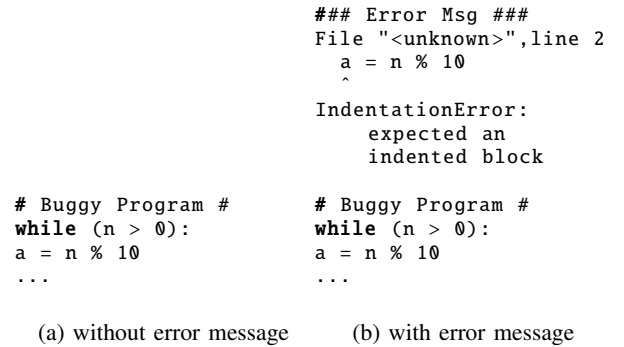

    \centering
    \begin{subfigure}[b]{.45\columnwidth}\centering
    \lstset{escapeinside={<@}{@>}}
	\begin{lstlisting}
# Buggy Program #
while (n > 0):
a = n % 10
...
\end{lstlisting}
\caption{without error message}
\label{subfig:motiv1_incorrect}
    \end{subfigure}%
        \begin{subfigure}[b]{.45\columnwidth}\centering
        \lstset{escapeinside={<@}{@>}}
	\begin{lstlisting}
### Error Msg ###
File "<unknown>",line 2
  a = n % 10
  ^
IndentationError: expected an indented block

# Buggy Program #
while (n > 0):
a = n % 10
...
\end{lstlisting}
\caption{with error message}
\label{subfig:motiv1_DebugS}
    \end{subfigure}%
    \caption{
    The syntax prompt generator produces
    prompts that can include the buggy program
    or the error message. We elide portions of the code fragments
    for brevity.}
    \label{fig:syntax-phase}
\end{figure}

The syntax prompt
generator produces two (multimodal) prompts, one with and one without the
syntax error message reported by the syntax oracle. An example of both is shown in Figure~\ref{fig:syntax-phase}. 
Because the syntax oracle is available, we do not need to choose a single prompt template
for all programs, but instead we query the LLMC with both prompts, extract the code portion from each generation, merge it into the original program by replacing the lines corresponding to the current program chunk, and then rely on the syntax oracle to filter out invalid repairs.

If a program candidate has no syntax errors,
it can move on to the semantic phase. If any syntax errors remain,
the syntax phase is repeated on this candidate program. This iteration allows
the repair of multiple, spatially-independent, syntax errors. 
For our evaluation, we allow this procedure to iterate at most two times to limit repair times.

\subsection{Semantic Phase}

After \system{} has generated syntactically valid candidate programs,
the repair procedure moves to a semantic repair phase. Intuitively,
this phase incorporates information that allows the LLMC to generate candidate
programs that satisfy the programming assignment task, as determined
by a semantic oracle. Following the approach of existing work
in automated repair for programming assignments~\cite{clara, Refactory}, 
we use the instructor's test suite (consisting of inputs and expected outputs)
as the semantic oracle. We say a program has been repaired if it 
produces the expected outputs for the given inputs.

\subsubsection{Semantic Prompt Generator}
The semantic prompt generator takes advantage of the rich set of signals
available in the education domain.
In particular, we exploit the fact that programming assignments typically
have available: (1) a natural language description of the task, 
(2) a set of test cases, and (3) peers' programming solutions.

The semantic prompt generator takes as input a syntactically
valid program, the task description in natural language, and the
set of instructor-provided test cases. The generator
then produces prompts with different combinations of this information.
Figure~\ref{fig:promptexample} shows an example of 
such a multimodal prompt. This prompt includes the student's buggy code, 
the natural language description of the assignment, as well
as the input-output-based test cases.

{
\begin{figure}[!t]
\begin{minipage}{1\columnwidth} 
\begin{lstlisting}[]
[[Buggy Program]]
### Buggy Program ###
x=input()
y=int(x)
z = number % 10
y = 10 * y + z
number = number / 10
number = int(number)
print("Reverse: {}".format(x[::-1]))
print("Sum: {}".format(Sum))


[[Problem Description]]
#Write a program to read a number (int) from the user. Print the number in reverse. Also print the sum of the number and its reverse in a separate line. See the examples.
#NOTE: Do not print any prompt in the input().

[[Test Suite]]
#input:
43
#output:
Reverse: 34
Sum: 77

#input:
500
#output:
Reverse: 5
Sum: 505

### Correct Program ###

\end{lstlisting}
\end{minipage}
    \caption{
    An example multimodal prompt (in zero-shot setting for brevity)
    produced by the semantic prompt generator.
     This prompt includes code, natural language, and test cases.
     Lines starting with the double brackets are shown only for clarity.
     }
    \label{fig:promptexample}
\end{figure}
}

If \system{} has access to other students' assignment solution history,
then it can also employ few-shot learning, described in the
following Section~\ref{sec:few-shot}, in each of these prompts.

Similarly to the syntax phase, rather than picking a single prompt template,
we use all prompts generated and rely on the semantic
oracle to identify
viable repair candidates. Each prompt given to the LLMC
can generate up to $K$ candidates, where we heuristically set $K$ to ten to balance exploration of candidates with search space explosion. 
Each of these candidates is given to the semantic oracle, which executes that candidate on the
test suite. We remove any candidate programs that result in 
a runtime exception or fail to satisfy any test cases.

If there are multiple valid candidate programs after the semantic phase,
we return the one with the smallest token-based 
edit distance~\cite{BIFI} 
to the student as the repaired program.

\subsubsection{Few-Shot Learning}\label{sec:few-shot}
If \system{} has access to other students' programs it can employ few-shot learning
In contrast to other repair systems, such as Refactory~\cite{Refactory}, that typically employ only correct
programs, \system{}'s few-shots consist of both correct \emph{and}
incorrect programs.

In particular, \system{}'s few-shot learning example bank consists of
pairs of program versions $(p, p')$ where both $p$ and $p'$
satisfy the syntax oracle, $p'$ satisfies the semantic oracle
but $p$ does not, and $p$ is a historical edit-version ancestor
of $p'$. Given a candidate program produced by 
the syntax phase of \system{}, we retrieve the
three most similar $p$ and their associated correct versions $p'$
to include as shots in the LLMC prompts produced by the
semantic prompt generator.

We take inspiration from traditional automated program repair
and say two programs are similar if they result in similar
test suite executions. We define a test suite execution vector
for program $p$ that captures test failures as
$T_p \in \mathbb{B}^n = (t_0, \cdots, t_n)$ where $n$ is the number of
test cases, and $t_i$ is the boolean failure status of the $i$th test. 
We define the similarity function between $p_1$ and $p_2$
as $1 - \textsc{Hamming}(T_{p_1}, T_{p_2})$ with \textsc{Hamming} the normalized Hamming distance~\cite{Hamming} between the two vectors.

Figure~\ref{fig:few-shots} shows an illustrative buggy student program, along with a single shot, consisting of the incorrect and corrected program from a peer.
The shot is picked as it has the same test execution vector as our target buggy program.

\begin{figure}[t]
    \centering
    
    \lstset{escapeinside={<@}{@>}}
	\begin{lstlisting}
[[Shot Starts]]
# Incorrect Program #
print (m+n)
# Correct Program #
print (m*n)
[Shot Ends]

[[Buggy Program Starts]] 
### Buggy Program ###
sum = m 
i = 0
while i < n:
  sum += 1
  i += 1
print (sum)
[[Buggy Program Ends]] 

[[Test Suite Starts]]
#input:
2 2
#output:
4

#input:
2 3
#output:
6
[[Test Suite Ends]]

### Correct Program ###


\end{lstlisting}
    \caption{
    An illustrative example of few-shot learning in \app.
    Both the incorrect program in the shot and the target buggy program to be repaired have the same test suite execution results [pass, fail].
    }
    \label{fig:few-shots}
\end{figure}

Note that if \system{} does not have access to peer programs,
then it can still query the LLMC using a zero-shot approach.
In our evaluation (Section~\ref{sec:eval}) we show that this ablated strategy still performs competitively.
\section{Evaluation}
\label{sec:eval}

We explore the following two research questions in our evaluation of \app:

\begin{itemize}
\item \textbf{(RQ1)} How does \app's overall repair rate (syntax and semantics)
compare to a state-of-the-art baseline?
\item \textbf{(RQ2)} What is the impact of 
the underlying design decisions
in \app? Specifically, what is the impact
of the structure-based program chunking, 
iterative querying, test-case-based few-shot selection, and multimodal ensembled prompts?

\end{itemize}

\para{Implementation.} We have built a \app prototype using a mix of Python and open-source software libraries. The core of \app's implementation consists of approximately 600 lines of Python code, which is 5 to 10 times less than a typical symbolic repair system in the education domain~\cite{clara,Refactory,rolim2017learning}.
In addition to the reduced engineering efforts, \app can handle both syntax and semantic bugs in one system, while most systems focus on one of the two bug classes.

We selected the top 10 program candidates in each syntax and semantics phase based on the 
on the average token 
log probabilities produced
by the LLMC.

For the model selection, we used OpenAI's Codex as our LLMC. Specifically, we used
the completion model. We found that other models, such as Codex Edit~\cite{codex-edit-api},
did not perform as well. We set the temperature to 0.8 based on preliminary experiments.

\para{Benchmark.} We derived a benchmark set by
selecting programs from a collection of
introductory Python assignments
collected by
third-party authors in a large Indian
university. This dataset is a Python-version
of the dataset described in~\cite{MACER}.

The dataset contains 18 assignments,
each with a problem description,
the test suite, and students' authoring history.
A student's history consists of an ordered collection of program versions, where each version can be an
explicit submission to the testing server,
or a periodic (passive) snapshot --
the dataset does not have a way to distinguish between these.

We removed three assignments that required reading files that are not reported in the dataset or that asked students to generate a PDF plot, which makes assessing
correctness difficult without manual inspection.
For each assignment, we selected the students that had an eventually correct program. For each such student, we collected the latest (closest to the correct version in time) version that had a syntactic mistake.  If the student never made a syntactic mistake, we remove this student from our benchmark. This is how we filtered programs containing syntax errors, and we expected the selected programs also have semantic errors that need to be repaired. This results in a total of 286 \emph{program pairs}, consisting of a buggy and ground-truth correct program version. 

\para{Baseline.} Most repair systems focus on either syntax repairs, or semantic repairs.\footnote{A notable exception in the education domain is sk\_p\cite{SKP}, however this tool is not publicly available and the repair rate (29\%) described in the paper is low compared to our baselines.}
To create a state-of-the-art baseline that
performs \emph{both} syntax and semantic repairs, we combined BIFI,
a state-of-the-art transformer-based Python syntax repair tool, and Refactory,
a state-of-the-art semantics repair tool
designed for introductory Python assignments.

To run this baseline, we gave BIFI the original student program with syntax errors and generated 50 candidate programs for each buggy program.
For each candidate, we ran the syntax oracle and checked for syntactic correctness.
For each candidate that passed the syntax check,
we called Refactory along with 
the instructor's reference solution.\footnote{The original Refactory paper shows that there is little-to-no performance difference between providing one and multiple correct reference programs.}
If Refactory can repair \emph{any}
of the candidates, we say it has repaired
the student's program. If there are multiple candidate programs that passed the test suite, we choose the one with the smallest token edit distance with respect to the buggy program as the final repair.
We ran all experiments on a Windows VM with an Intel i7 CPU and 32 GB of RAM.

\label{subsec:analysis}

\setlength{\tabcolsep}{.8em}
\def\arraystretch{1.2}
\begin{table*}[t]
\centering
\caption{
\system{} (without few shots) repairs
a larger fraction of programs (86.71\%) compared
to our baseline (67.13\%). On average, 
\system{}
repairs are closer in terms of
token edit distance (TED) to the original student program
(31.40 versus 42.50). Adding few-shots based on
other peers' programs raises \system{}'s
repair rate to 96.50\% while keeping a comparable
average token edit distance (31.29).
}
\label{table:res_union_iterative_fewshot}
\begin{footnotesize}
\resizebox{2.05\columnwidth}{!}{
\begin{tabular}{|cc|cc|cc|cc|}
\hline
\multicolumn{2}{|c|}{Method}                                      & \multicolumn{2}{c|}{\app (without few-shot learning)}                  & \multicolumn{2}{c|}{\app (with few-shot learning)}      & \multicolumn{2}{c|}{BIFI + Refactory}      \\ \hline
\multicolumn{1}{|c|}{Problem ID} & \# Submissions & \multicolumn{1}{c|}{Repair rate (\%)} & Mean TED (SD)    & \multicolumn{1}{c|}{Repair rate (\%)} & Mean TED (SD)    & \multicolumn{1}{c|}{Repair rate (\%)} & Mean TED (SD)    \\ \hline
\multicolumn{1}{|c|}{2865}    & 11   & \multicolumn{1}{c|}{100.00}                    & 6.45 (4.74)      & \multicolumn{1}{c|}{100.00}                    & 6.45 (4.74)    & \multicolumn{1}{c|}{100.00}                      & 5.28 (4.27)  \\ \hline
\multicolumn{1}{|c|}{2868}    & 28   & \multicolumn{1}{c|}{85.71}                     & 8.79 (8.94)      & \multicolumn{1}{c|}{100.00}                    & 8.64 (8.49)    & \multicolumn{1}{c|}{82.14}                    & 8.35 (7.00) \\ \hline
\multicolumn{1}{|c|}{2869}    & 23   & \multicolumn{1}{c|}{95.65}                     & 16.68 (18.47)    & \multicolumn{1}{c|}{100.00}                    & 10.30 (10.99)  & \multicolumn{1}{c|}{69.57}                    & 17.81 (13.67) \\ \hline
\multicolumn{1}{|c|}{2870}    & 27   & \multicolumn{1}{c|}{74.07}                     & 10.00 (13.33)    & \multicolumn{1}{c|}{100.00}                    & 15.00 (19.35)  & \multicolumn{1}{c|}{85.19}                    & 15.74 (23.92) \\ \hline
\multicolumn{1}{|c|}{2872}    & 18   & \multicolumn{1}{c|}{100.00}                    & 8.33 (15.15)     & \multicolumn{1}{c|}{100.00}                    & 7.39 (13.01)   & \multicolumn{1}{c|}{72.22}                    & 76.61 (9.24)\\ \hline
\multicolumn{1}{|c|}{2873}    & 32   & \multicolumn{1}{c|}{78.13}                     & 12.00 (16.18)    & \multicolumn{1}{c|}{90.63}                     & 12.93 (15.47)  & \multicolumn{1}{c|}{84.38}                    & 58.48 (14.04)\\ \hline
\multicolumn{1}{|c|}{2874}    & 16   & \multicolumn{1}{c|}{100.00}                    & 9.56 (12.50)     & \multicolumn{1}{c|}{100.00}                    & 8.50 (11.76)   & \multicolumn{1}{c|}{87.50}                    & 23.43 (16.28) \\ \hline
\multicolumn{1}{|c|}{2875}    & 23   & \multicolumn{1}{c|}{86.96}                     & 14.75 (19.97)    & \multicolumn{1}{c|}{100.00}                    & 11.52 (12.52)  & \multicolumn{1}{c|}{78.26}                    & 35.72 (14.58)\\ \hline
\multicolumn{1}{|c|}{2877}    & 21   & \multicolumn{1}{c|}{100.00}                    & 9.71 (16.82)     & \multicolumn{1}{c|}{100.00}                    & 9.14  (16.79)  & \multicolumn{1}{c|}{80.95}                    & 33.12 (24.56) \\ \hline
\multicolumn{1}{|c|}{2878}    & 25   & \multicolumn{1}{c|}{100.00}                    & 37.00 (60.16)    & \multicolumn{1}{c|}{100.00}                    & 36.32 (59.53)  & \multicolumn{1}{c|}{68.00}                    & 97.94 (23.79) \\ \hline
\multicolumn{1}{|c|}{2879}    & 21   & \multicolumn{1}{c|}{76.19}                     & 131.19 (51.62)   & \multicolumn{1}{c|}{85.71}                     & 132.78 (52.61) & \multicolumn{1}{c|}{52.38}                    & 125.64 (11.33) \\ \hline
\multicolumn{1}{|c|}{2882}    & 23   & \multicolumn{1}{c|}{60.87}                     & 90.64 (71.76)    & \multicolumn{1}{c|}{91.30}                     & 106.57 (77.57) & \multicolumn{1}{c|}{0.00}                     & N/A \\ \hline
\multicolumn{1}{|c|}{2883}    & 5    & \multicolumn{1}{c|}{100.00}                    & 17.40 (14.67)    & \multicolumn{1}{c|}{100.00}                    & 17.40 (14.67)  & \multicolumn{1}{c|}{40.00}                    & 53 (0.00) \\ \hline
\multicolumn{1}{|c|}{2920}    & 10   & \multicolumn{1}{c|}{80.00}                     & 84.38 (67.62)    & \multicolumn{1}{c|}{80.00}                     & 53.50 (66.05)  & \multicolumn{1}{c|}{0.00}                     & N/A    \\ \hline
\multicolumn{1}{|c|}{2921}    & 3    & \multicolumn{1}{c|}{100.00}                    & 28.00 (3.61)     & \multicolumn{1}{c|}{100.00}                    & 28.00 (3.61)   & \multicolumn{1}{c|}{0.00}                     & N/A     \\ \hline
\hline
  \multicolumn{2}{|c|}{Overall}       & \multicolumn{1}{c|}{86.71}                    &   31.40   & \multicolumn{1}{c|}{96.50}                    & 31.29  & \multicolumn{1}{c|}{67.13}                     &   42.50   \\ \hline
\end{tabular}
}
\end{footnotesize}
\end{table*}

\subsection{RQ1: Overall Repair Performance}
Table~\ref{table:res_union_iterative_fewshot} shows that without few-shot learning
\app can repair 86.71\% of student programs.
In contrast, our baseline repairs
67.13\% of student programs. In addition,
\app repairs are closer to the original
student programs. In particular, the mean token edit distance between the buggy program and our repaired program is 31.40 compared to 42.50 for our baseline.

Our results also show that if we add
few-shot learning, by leveraging the
availability of other student's
incorrect and correct versions paired with
our test-case-based few-shot selection strategy,
we can raise \app's overall repair rate to 96.50\%.
It is worth noting that the increase
in repairs does not raise average token edit distance
(31.29). In this setting, \app outperformed the baseline by 29.37 percentage points
in terms of repair rates and reduced average token edit distance
by 26.38\%.

A key idea behind Refactory is to perform a control-flow match between
correct programs and the student's buggy program. To do this, Refactory
generates multiple versions of the correct program (in our case, the reference
solution) by applying rewrites. Repairs that require substantial refactoring,
such as that shown in the motivating example (Section~\ref{sec:motiv}) are 
beyond Refactory's expressiveness.

\setlength{\tabcolsep}{.8em}
\def\arraystretch{1.2}
\begin{table}[t]
\caption{
The first stage in the repair process is to fix
syntax errors. \system{} can produce
a syntactically valid candidate for all programs
in our benchmark, compared to 80.07\% for BIFI. 
On average, \system{}'s repairs are also closer to the original program (edit distance of 5.46 versus 25.07).
}
\label{table:res_union_chunk_bifi}
\begin{footnotesize}
\resizebox{1.0\columnwidth}{!}{
\begin{tabular}{|cc|c|cc|}
\hline
\multicolumn{2}{|c|}{Method}                     & \app   & \multicolumn{2}{|c|}{BIFI} \\ \hline
\multicolumn{1}{|c|}{ID} & \# Sub & Mean TED (SD) & \multicolumn{1}{|c|}{Repair rate (\%)}   & Mean TED (SD)   \\ \hline
\multicolumn{1}{|c|}{2865}       & 11             & 2.18 (1.25)           & \multicolumn{1}{|c|}{100.00}    & 1.82 (0.75)      \\ \hline
\multicolumn{1}{|c|}{2868}       & 28                & 2.75 (2.17)             & \multicolumn{1}{|c|}{82.14}    & 1.83 (1.11)       \\ \hline
\multicolumn{1}{|c|}{2869}       & 23               & 2.91 (2.41)           & \multicolumn{1}{|c|}{73.91}  & 1.47 (0.80)        \\ \hline
\multicolumn{1}{|c|}{2870}       & 27              & 2.33 (2.18)          & \multicolumn{1}{|c|}{85.19}    & 2.04 (1.89)      \\ \hline
\multicolumn{1}{|c|}{2872}       & 18                 & 2.39 (1.2)           & \multicolumn{1}{|c|}{72.22}     & 1.62 (0.87)         \\ \hline
\multicolumn{1}{|c|}{2873}       & 32            & 2.84 (2.58)          & \multicolumn{1}{|c|}{84.38}    & 2.56 (2.04)     \\ \hline
\multicolumn{1}{|c|}{2874}       & 16                & 2.06 (1.84)           & \multicolumn{1}{|c|}{87.50}    & 2.07 (2.02)       \\ \hline
\multicolumn{1}{|c|}{2875}       & 23              & 2.78 (2.71)            & \multicolumn{1}{|c|}{78.26}   & 1.78 (1.56)    \\ \hline
\multicolumn{1}{|c|}{2877}       & 21             & 2.19 (1.29)            & \multicolumn{1}{|c|}{80.95}   & 3.18 (7.47)         \\ \hline
\multicolumn{1}{|c|}{2878}       & 25             & 4.84 (8.58)            & \multicolumn{1}{|c|}{0.00}  & 40.2 (60.65)     \\ \hline
\multicolumn{1}{|c|}{2879}       & 21             & 18.86 (21.24)         & \multicolumn{1}{|c|}{66.67}   & 117.00 (58.15)      \\ \hline
\multicolumn{1}{|c|}{2882}       & 23             & 17.39 (23.23)        & \multicolumn{1}{|c|}{86.96}    & 127.65 (74.76)      \\ \hline
\multicolumn{1}{|c|}{2883}       & 5               & 5.60 (9.74)           & \multicolumn{1}{|c|}{80.00}      & 36.25 (37.98)       \\ \hline
\multicolumn{1}{|c|}{2920}       & 10             & 10.30 (18.68)              & \multicolumn{1}{|c|}{50.00}   & 51.75 (54.90)     \\ \hline
\multicolumn{1}{|c|}{2921}       & 3              & 1.67 (0.58)             & \multicolumn{1}{|c|}{100.00}     & 1.33 (0.58)      \\ \hline
\hline
\multicolumn{2}{|c|}{Overall}            & 5.46          & \multicolumn{1}{|c|}{80.07}     & 25.07     \\ \hline
\end{tabular}
}
\end{footnotesize}
\end{table}

Repairing semantic mistakes typically depends on first resolving any 
syntactic mistakes. Indeed, students often focus on resolving mistakes reported
by the parser/compiler before they move on to debugging test-case failures.
\app's architecture, as well as our composed baseline, reflects this approach.
As a result, we also want to understand syntax repair performance.

Table~\ref{table:res_union_chunk_bifi} summarizes the \emph{syntax repair rates}
across assignments and approaches. Our results show
that \app repairs the syntax bugs in all of the 286 programs, with a 100\% syntax repair rate. 
This outperforms the state-of-the-art BIFI, which has a syntax repair rate of 80.07\%.
In addition, \app's repairs have a 
substantially lower mean token edit distance (5.46 versus 25.07), meaning
our repairs on average introduce fewer changes to the original programs, which may facilitate
understanding of the fixes.

We also observed that in 17 out of 286 cases, BIFI fails to handle the input
program, potentially due to lexer issues. This highlights another advantage of using \app to repair programs because \app does not have any constraints over the input as a result of its prompt-based learning strategy.

BIFI is very effective at repairing small syntax mistakes
in assignments of lower difficulty. For example, in assignment 2865, BIFI repairs
all syntax errors and does so with a smaller average token edit distance (1.82 versus 2.18) compared to \app. One interesting direction for future work is to combine BIFI with \app, as the repairs can be complementary. In this case, Codex could focus on generating more complex repairs and BIFI could focus on small edits for simpler tasks such as missing a quote in a string.

\subsection{RQ2: Ablation Study}

We now present the results of experiments to analyze different design choices in \app. 
\app uses multimodal prompts, iterative querying, test-case-based few-shot selection, and structure-based program chunking to repair student mistakes.
The power of few-shot selection was already shown in Table~\ref{table:res_union_iterative_fewshot}. We will now present the results of the other three design choices.

\subsubsection{Program Chunking}
In the syntax stage, \app first extracts program chunks from the original
buggy program as detailed in Section~\ref{sec:method}. The intuition is that these chunks contain the syntax error
we want to fix, along with surrounding context, while excluding code lines
that are not relevant to the fix. Our goal is to reduce the number of (spurious) edits produced by the LLMC by reducing the code surface in the prompt.

To evaluate the impact of program chunking, we removed it from \app and compared
performance to the original \app.
Table~\ref{table:chunk} shows the average token edit distance produced
in the syntax phase with and without program chunking. We found
that program chunking can reduce the average token edit distance up to 56.32\% 
(problem assignment 2878). Overall, the average token edit distance is reduced
from 9.38 to 5.46 (41.79\%) by adding program chunking.

\setlength{\tabcolsep}{.8em}
\def\arraystretch{1.2}
\begin{table}[t]
\caption{
Programs chunking reduces the average token edit distance
across all assignments. PG is short for performance gain.}
\label{table:chunk}
\begin{footnotesize}
\resizebox{1\columnwidth}{!}{
\begin{tabular}{|c|c|c|c|}
\hline
\multicolumn{1}{|c|}{Method}                     & \multicolumn{1}{|c|}{\app (no chunking)}        & \multicolumn{1}{|c|}{\app (with chunking)} &  \\ \hline
\multicolumn{1}{|c|}{ID} & Mean TED (SD) & Mean TED (SD)   & PG (\%)   \\ \hline
\multicolumn{1}{|c|}{2865}                  & 2.45 (1.21)    & 2.18 (1.25)      & 11.11          \\ \hline
\multicolumn{1}{|c|}{2868}                  & 2.82 (2.14)    & 2.75 (2.17)      & 2.53             \\ \hline
\multicolumn{1}{|c|}{2869}                  & 2.91 (2.41)    & 2.91 (2.41)      & 0.00             \\ \hline
\multicolumn{1}{|c|}{2870}                   & 2.33 (2.18)    & 2.33 (2.18)      & 0.00            \\ \hline
\multicolumn{1}{|c|}{2872}                & 2.44 (1.2)     & 2.39 (1.2)       & 2.27             \\ \hline
\multicolumn{1}{|c|}{2873}                & 3.09 (2.61)    & 2.84 (2.58)      & 8.08            \\ \hline
\multicolumn{1}{|c|}{2874}                 & 2.25 (2.08)    & 2.06 (1.84)      & 8.33              \\ \hline
\multicolumn{1}{|c|}{2875}                  & 3.52 (4.13)    & 2.78 (2.71)      & 20.99          \\ \hline
\multicolumn{1}{|c|}{2877}                  & 2.29 (1.27)    & 2.19 (1.29)      & 4.17               \\ \hline
\multicolumn{1}{|c|}{2878}                   & 11.08 (20.3)   & 4.84 (8.58)      & 56.32             \\ \hline
\multicolumn{1}{|c|}{2879}                  & 33.14 (24.91)  & 18.86 (21.24)    & 43.10            \\ \hline
\multicolumn{1}{|c|}{2882}                   & 42.57 (41.54)  & 17.39 (23.23)    & 59.14            \\ \hline
\multicolumn{1}{|c|}{2883}                  & 6.20 (11.08)   & 5.60 (9.74)      & 9.68            \\ \hline
\multicolumn{1}{|c|}{2920}                   & 15.20 (19.45)  & 10.30 (18.68)    & 32.24            \\ \hline
\multicolumn{1}{|c|}{2921}                 & 1.67 (0.58)    & 1.67 (0.58)      & 0.00            \\ \hline
\hline
\multicolumn{1}{|c|}{Overall}           & 9.38    & 5.46      & 41.79       \\ \hline
\end{tabular}
}
\end{footnotesize}
\end{table}

\subsubsection{Iterative Querying}
Students typically resolve syntax errors first and then move on to resolving
semantic mistakes. \app's architecture follows
this same intuition. To compare the effectiveness of this iterative approach,
we ran a variant of \app that addresses both syntax and semantic bugs in a \emph{single}
round. Table~\ref{table:iterative} shows the results of this ablated
variant and full \app (without few-shots). We find that splitting concerns into two phases results
in an increase in the overall repair rate from 82.87\% to 86.71\%. Using two phases
increases the average token edit distance slightly (30.29 to 31.40).

\setlength{\tabcolsep}{.8em}
\def\arraystretch{1.2}
\begin{table}[t]
\centering
\caption{
\system{} performs iterative querying, splitting
the repair procedure into a syntax and a semantic phase.
We find that this iterative approach raises
the overall repair rate from 82.87\% to
86.71\% (without few-shots).
RR stands for repair rate.
}
\label{table:iterative}
\begin{footnotesize}
\resizebox{1.0\columnwidth}{!}{
\begin{tabular}{|c|cc|cc|}
\hline
\multicolumn{1}{|c|}{Method}                                      & \multicolumn{2}{c|}{\app (no iterative)}                  & \multicolumn{2}{c|}{\app (with iterative)}          \\ \hline
\multicolumn{1}{|c|}{ID} & \multicolumn{1}{c|}{RR (\%)} & Mean TED (SD)    & \multicolumn{1}{c|}{RR (\%)} & Mean TED (SD)   \\ \hline
\multicolumn{1}{|c|}{2865}    & \multicolumn{1}{c|}{100.00}                    & 6.45 (4.74)  & \multicolumn{1}{c|}{100.00}                    & 6.45 (4.74)       \\ \hline
\multicolumn{1}{|c|}{2868}    & \multicolumn{1}{c|}{85.71}                     & 8.92 (8.88)  & \multicolumn{1}{c|}{85.71}                     & 8.79 (8.94)    \\ \hline
\multicolumn{1}{|c|}{2869}     & \multicolumn{1}{c|}{86.96}                     & 13.35 (12.36) & \multicolumn{1}{c|}{95.65}                     & 16.68 (18.47)   \\ \hline
\multicolumn{1}{|c|}{2870}      & \multicolumn{1}{c|}{70.37}                     & 11.42 (13.87)  & \multicolumn{1}{c|}{74.07}                     & 10.00 (13.33) \\ \hline
\multicolumn{1}{|c|}{2872}       & \multicolumn{1}{c|}{100.00}                    & 8.50 (15.22)  & \multicolumn{1}{c|}{100.00}                    & 8.33 (15.15)    \\ \hline
\multicolumn{1}{|c|}{2873}    & \multicolumn{1}{c|}{71.88}                     & 9.48 (11.63) & \multicolumn{1}{c|}{78.13}                     & 12.00 (16.18)  \\ \hline
\multicolumn{1}{|c|}{2874}    & \multicolumn{1}{c|}{100.00}                    & 9.75 (12.51) & \multicolumn{1}{c|}{100.00}                    & 9.56 (12.50)      \\ \hline
\multicolumn{1}{|c|}{2875}   & \multicolumn{1}{c|}{82.61}                     & 13.16 (18.69)  & \multicolumn{1}{c|}{86.96}                     & 14.75 (19.97)   \\ \hline
\multicolumn{1}{|c|}{2877}     & \multicolumn{1}{c|}{100.00}                    & 9.71 (16.82)   & \multicolumn{1}{c|}{100.00}                    & 9.71 (16.82)     \\ \hline
\multicolumn{1}{|c|}{2878}      & \multicolumn{1}{c|}{100.00}                    & 38.16 (62.24) & \multicolumn{1}{c|}{100.00}                    & 37.00 (60.16)   \\ \hline
\multicolumn{1}{|c|}{2879}     & \multicolumn{1}{c|}{71.43}                     & 130.07 (53.23)   & \multicolumn{1}{c|}{76.19}                     & 131.19 (51.62)  \\ \hline
\multicolumn{1}{|c|}{2882}    & \multicolumn{1}{c|}{56.52}                     & 97.85  (72.64)   & \multicolumn{1}{c|}{60.87}                     & 90.64 (71.76)   \\ \hline
\multicolumn{1}{|c|}{2883}   & \multicolumn{1}{c|}{100.00}                    & 17.40 (14.67)  & \multicolumn{1}{c|}{100.00}                    & 17.40 (14.67) \\ \hline
\multicolumn{1}{|c|}{2920}       & \multicolumn{1}{c|}{50.00}                     & 50.20 (48.9) & \multicolumn{1}{c|}{80.00}                     & 84.38 (67.62)    \\ \hline
\multicolumn{1}{|c|}{2921}     & \multicolumn{1}{c|}{100.00}                    & 28.00 (3.61) & \multicolumn{1}{c|}{100.00}                    & 28.00 (3.61)     \\ \hline
\hline
  Overall       & \multicolumn{1}{c|}{82.87}                    &   30.29   & \multicolumn{1}{c|}{86.71}                    & 31.40    \\ \hline
\end{tabular}
}
\end{footnotesize}
\end{table}

\subsubsection{Multimodal Prompts}
\app combines different types of input (code, natural language, test cases)
into its prompts. This richness of inputs is a particular advantage of the
educational setting. \app
\emph{ensembles} these various prompts by querying the LLMC and \emph{then}
relying on the (syntax or semantics) oracle to filter out candidates. 
This approach is based on the idea that different prompts may produce complementary
candidates. To demonstrate this complementarity, we generate different prompt structures, use them in a single query to the LLMC, and evaluate the overall repair rate across assignments.

\begin{figure}[t]
  \centering
  \includegraphics[width=1.0\linewidth]{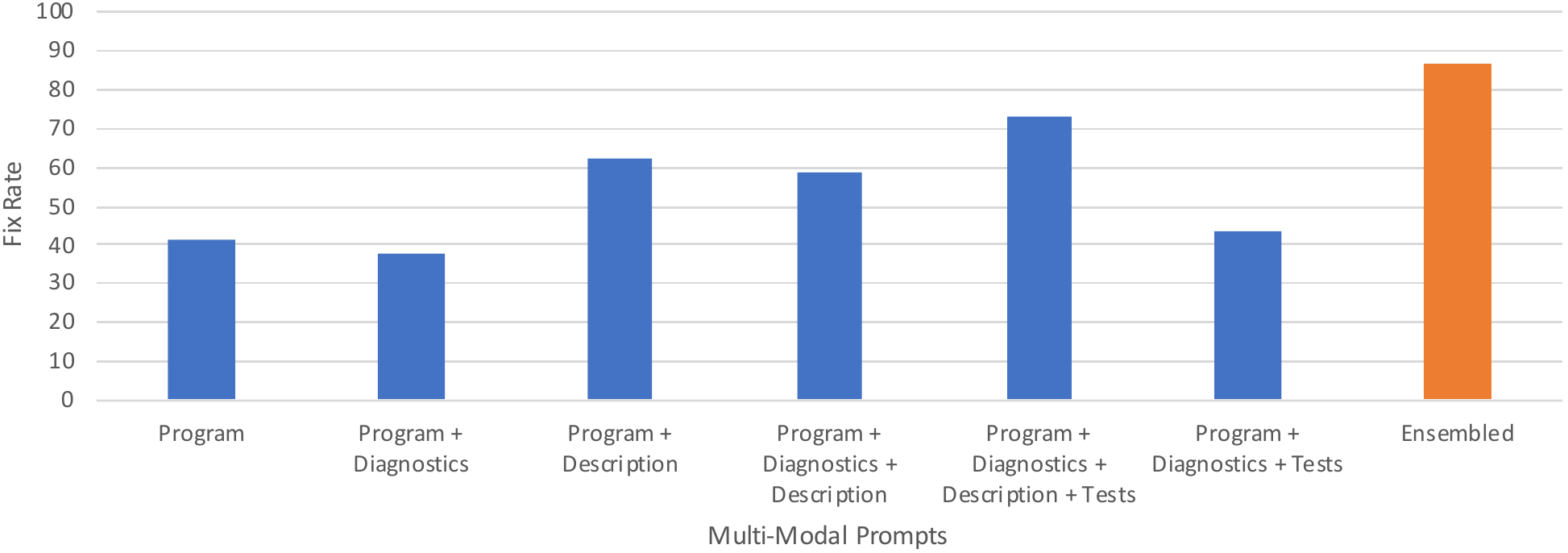}
  \caption{
  Rather than pick a single prompt
  structure, \system{} \emph{ensembles} them
  by querying the LLMC with multiple prompts,
  and then relying on the (syntax and semantic) oracles
  to rule out invalid candidates. Ensembling
  outperforms any particular prompt structure 
  as candidates are complementary.
  }
  \label{fig:comp_semantic}
\end{figure}

Figure~\ref{fig:comp_semantic} shows that different prompt structures result
in different overall performances in terms of fix rate. If a single prompt structure
needs to be chosen, \lstinline{Program + Diagnostics + Description + Tests} structure
is most effective in this experiment. However, if we \emph{ensemble} the candidates,
we obtain the best result as candidates are complementary.

\subsubsection{Combining BIFI and Refactory to Repair Programs}
We investigated why our baseline produces
repairs that (on average) have a larger token
edit distance compared to \system{}.
We found that in some cases, BIFI produces repairs by deleting a portion of the code snippet that contains the syntax errors.
Although this is an effective way to deal with syntax errors, it makes repairing semantic errors harder by deleting parts that may capture crucial logic in the original source code.

Below is one such example from our evaluation. The code snippet contains a syntax mistake in the last line.
The parser complains that the \textit{``Expression cannot contain an assignment =''}.
In particular, the student has written an 
equals when they should have used a plus operator (which corresponds to the repair produced by \system{}).

\lstset{
    language=Python,
    numbers=left,
    stepnumber=1,
    escapeinside={<@}{@>}
}

\begin{minipage}{.9\columnwidth}
\begin{lstlisting}[xleftmargin=.03\textwidth]
marksSum={}
for i in total:
  if int(i[0])not in marksSum:
    marksSum[int(i[0])]= int(i[2])
  else:
    k=int(i[2])
    marksSum[int(i[0])]+=k
for i in sorted(marksSum):
  print(str(i)<@\textcolor{red}{=}@>":"+str(marksSum[i]))
\end{lstlisting}
\end{minipage}

However, BIFI produced a different fix by
removing the second \lstinline{for} loop (lines 8-9) completely.
This deletion introduced challenges for Refactory in the later semantic repair phase.
Although Refactory in the end successfully repaired this program, the repair it generated is syntactically equivalent to the reference solution and is effectively completely rewritten with respect to the original incorrect program.

Overall, our comparison between \app and BIFI + Refactory highlights the
challenges in combining the state-of-the-art syntax and semantics tools to repair incorrect introductory programming assignments.
BIFI and Refactory each focus on their targets, syntax bugs repair and semantic bugs repair, respectively.
Combining them does not guarantee the best overall repair (as in the example shown above).
Additionally, combining BIFI and Refactory required non-trivial engineering efforts, further motivating the 
need for a unified approach that can handle both types of bugs.
\section{Threats to Validity}
\app validates candidate repairs by comparing execution results on the test suite with the reference program given by instructors.
Validating program correctness through tests is not as strong as formal verification.
To the best of our knowledge, the use of tests as a proxy for correctness is standard in the educational domain~\cite{Autograder,clara}.

We carried out our evaluation on one particular set of 286 student program. The size of the dataset is on par with the state-of-the-art automated program repair techniques\cite{verifix,Li22generating}, but
increasing the size of the evaluation
dataset may provide additional insights and presents an opportunity for future work.
We only evaluated \system{} on Python programs. However, our prompt-based approach can be applied to other languages.
\section{Related Work}\label{sec:related}

\para{Automated Program Repair.}
The software engineering and programming languages community
has a long history of developing tools for automatically
repairing buggy programs. Existing approaches have applied
a variety of technical ideas, including 
formal program logic~\cite{Angelix,MechtaevNNGR18,CPR}, search-based
techniques~\cite{wong21varfix} like genetic programming~\cite{GenProg,RandomQi,PAR},
and machine learning~\cite{prophet,genesis,wang2018dynamic}. A particularly popular approach
to APR consists of generating many program candidates,
typically derived by performing syntactic transformations
of the original buggy program, and then validating these candidates
using test suite as an oracle~\cite{GenProg}.

Similarly, \app{} aims to automatically fix syntax and semantic
errors in buggy programs, and uses a syntax oracle (the compiler)
and semantic oracle (test cases) to validate candidate programs produced.
However, in contrast to existing work, \app{} employs a large language model
(Codex) as the main program transformation module and uses an ensemble of
multi-modal prompts to improve its success rate. Additionally, students struggle even with basic syntax errors~\cite{Altadmri15,Drosos17}. Therefore, \app{}
targets students' incorrect submissions, rather than 
professional developers' production bugs. As a result, \app{} has an additional
requirement of minimizing the size of the change made
to allow students to better learn from the repaired program.

\para{AI for Programming Education.}
AI, both symbolic and neural, has been extensively applied to the
domain of education~\cite{alphacode,Finnie2022}. In particular, for programming education past
research has explored topics including
feedback generation~\cite{Autograder,rolim2017learning,clara,SAR,Refactory,Cafe,jialu22} and program repair~\cite{YiFSE17,QiLeveraging17,wang2018dynamic,DinellaDLNSW20,BIFI,FAPR}.
\app{} is complementary to the series of work, showing that the task of program 
repair in this domain can be successfully tackled using an LLMC. Using such an approach can lower the
effort to develop and maintain an APR tool for introductory programming assignments.

\para{LLMs for Code Intelligence.}
Large pre-trained language models, such as OpenAI's Codex, 
Salesforce CodeGen~\cite{CodeGen}, and BigScience's BLOOM\cite{BLOOM},
have been shown to be effective for a range of code intelligence tasks.
For example, Microsoft's Copilot\cite{copilot} builds on Codex to produce
more effective single-line and multi-line code completion suggestions.
Prior work has shown that such LLMs can also be used for 
repairing programs outside of the educational context~\cite{VerbruggenLG21,RahmaniRGLMRST21,DinellaRML22,ZhangMKPL22}.
Using these models to perform code generation from informal specifications, such as natural language,
has also been a topic of active research~\cite{alphacode}.
Similarly to this work, \app{} uses Codex to perform a code intelligence task 
(program repair). However, \app{} is designed to focus on student programming
tasks and as such our design decisions (e.g., strategies to reduce token
edit distance) may not apply to other domains such as professional developers.
\section{Conclusion}\label{sec:conclusion}
In this work, we introduced an approach to 
repair syntactic and semantic mistakes
in introductory Python assignments.
At the core of our approach sits a 
large language model trained on code. 
We leverage multimodal
prompts, iterative querying, test-case-based
few-shot selection, and program chunking to 
produce repairs. We implement our approach
using Codex in a system called \system{}
and evaluate it on real student programs.
Our results show that our unified system
\system{} outperforms
a baseline, a combination of a state-of-the-art syntax
repair engine and a state-of-the-art semantic
repair engine, while producing smaller patches.

\bibliography{intern}
\bibliographystyle{IEEEtran}
\end{document}